\documentstyle[preprint,tighten,prl,aps]{revtex}
\begin{document}


\preprint{ITP-SB-95-06}
\title{Exact ground states of generalized Hubbard models}

\author{Jan de Boer\footnote{email:
{\sc deboer@insti.physics.sunysb.edu}}}
\address{Institute for Theoretical Physics\\
State University of New York at Stony Brook\\
Stony Brook, NY 11794-3840, U.S.A.}
\author{Andreas Schadschneider\footnote{email:
{\sc as@thp.uni-koeln.de}}
}
\address{Institut f\"ur Theoretische Physik\\
Universit\"at zu K\"oln\\
Z\"ulpicher Strasse 77\\
D-50937 K\"oln, Germany }
\date{\today}
\maketitle
\begin{abstract}
We present a simple method for the construction of exact ground states
of generalized Hubbard models in arbitrary dimensions. This method is
used to derive rigorous criteria for the stability of various ground
state types, like the $\eta$-pairing state, or N\'eel and
ferromagnetic states. Although the approach presented here is much
simpler than the ones commonly used, it yields better bounds for the
region of stability.
\end{abstract}
\pacs{75.10.Lp, 74.20-Mn, 71.20.Ad}

\narrowtext

In recent years there has been an increasing interest in analyzing
systems of correlated fermions by constructing the ground state
explicitly in certain parameter regimes, but arbitrary dimensions
\cite{BG}-\cite{montorsi}. In general those ground states have a
rather simple structure. Typically one can speak of quasi-classical
states where quantum fluctuations are not important and there are no
finite-size corrections to the ground state energy. E.g.\ the fully
polarized ferromagnetic state has been in the center of attention. In
\cite{dBKS} the corresponding program for superconducting states of
the $\eta$-pairing type \cite{pairing,dBKS} has been initiated. These
$\eta$-pairing states show Off-diagonal-Long-Range-Order (ODLRO)
\cite{odlro} which in turn implies the Meissner effect and flux
quantization \cite{odlro,sewell,nieh}, i.e.\ superconductivity.

The most popular method for the construction of exact ground states
was introduced by Brandt and Giesekus \cite{BG} and generalized by
Strack and Vollhardt \cite{strack1}. The basic idea is to start with
a Hamiltonian ${\cal H}_0$ with a known ground state
$|\,\psi_N\,\rangle$ and add operators ${\cal P}^\dagger {\cal P}$
where ${\cal P}$ annihilates $|\,\psi_N\,\rangle$. Thus ${\cal H}_0+
{\cal P}^\dagger {\cal P}$ also has $|\,\psi_N\,\rangle$ as ground
state. It is obvious that for a given Hamiltonian ${\cal H}$ and
ground state $|\,\psi_N\,\rangle$ it is usually not easy to find an
appropriate ${\cal H}_0$ and ${\cal P}$. Application of this method
requires a lot of guesswork and is not very systematic. In addition,
one does not gain much insight into the underlying physics by this
sort of construction.

Ovchinnikov \cite{ovchin,ovchin2} has used a different method to
improve some of the results obtained previously. His approach is based
on Gerschgorin's Theorem (see e.g.\ \cite{matrix}) which gives a lower
bound on the ground state energy of the Hamiltonian and thus
complements the usual variational principle which gives upper bounds.
If those bounds coincide each eigenstate with the corresponding energy
is a ground state.

In this paper we apply a much simpler and clearer method for the
construction of exact ground states which we call {\em Optimal Ground
State Approach} (OGS approach). The basic idea is to diagonalize the
local interaction and make all local eigenstates which are needed for
the construction of the global ground state $|\,\psi_N\,\rangle$ local
{\em ground} states. This usually implies some restrictions on the
interaction parameters, typically in the form of inequalities. We will
show on a few explicit examples that this method is not only much
simpler than the ones described above, but also reproduces or improves
the results for all cases considered previously. We like to point out
that the OGS method does not only work efficiently for states
$|\,\psi_N\,\rangle$ which are simple tensor products, but also for
more complicated ones, like the $\eta$-pairing states or so-called
matrix-product ground states of certain spin-1 chains \cite{KSZ}.
A more detailed exposition of the OGS
approach and a thorough comparision with the other methods will be
presented in a future publication \cite{dbs}.

For any Hamiltonian ${\cal H}=\sum_{\langle jl\rangle} h_{jl}$ on a
lattice with $L$ sites ($\langle jl \rangle$ denotes neighboring
sites) it is usually quite easy to diagonalize the local interaction
$h_{jl}$.  Suppose now that the lowest eigenvalue of $h_{jl}$ is $0$
(e.g.\ by adding a suitable constant). Clearly $0$ is a lower bound
for the global ground state energy. The global ground state
$|\,\psi_N\,\rangle$ is called {\em optimal} \cite{zitt} iff ${\cal
H}|\,\psi_N\,\rangle =0$, i.e.\ the ground state energy is just the
lower bound found by diagonalizing the local interaction.

In the following we will be interested in a generalized Hubbard model
with
\begin{eqnarray}
h_{jl} & = & -t \sum_{\sigma} ( c^{\dagger}_{j\sigma} c_{l\sigma} +
  c^{\dagger}_{l\sigma} c_{j\sigma} )
  +X\sum_{\sigma}( c^{\dagger}_{j\sigma} c_{l\sigma}
  + c^{\dagger}_{l\sigma} c_{j\sigma} ) (n_{j,-\sigma}+n_{l,-\sigma})
  \nonumber \\[2mm]
& & +\frac{U}{Z}\left( (n_{j\uparrow}-{\frac{1}{2}})(n_{j\downarrow}
  -{\frac{1}{2}})+ (n_{l\uparrow}-{\frac{1}{2}})(n_{l\downarrow}
  -{\frac{1}{2}})\right)\nonumber \\[2mm]
& & +V (n_j-1)(n_l-1)+Y \left(c^{\dagger}_{j\uparrow}
  c^{\dagger}_{j\downarrow} c_{l\downarrow} c_{l\uparrow}
  + c^{\dagger}_{l\uparrow} c^{\dagger}_{l\downarrow}
  c_{j\downarrow} c_{j\uparrow}\right)  \nonumber \\[2mm]
& & +\frac{J_{xy}}{2} \left(S^{+}_j S^{-}_l + S^{+}_l S^{-}_j\right)
    + J_z S^z_j S^z_l
+\frac{\mu}{Z}\left(n_j + n_l\right)\ .
\label{hloc1}
\end{eqnarray}
Here $c_{j\sigma}$ and $c^\dagger_{j\sigma}$ are the canonical Fermi
operators, $n_{j\sigma}=c^\dagger_{j\sigma}c_{j\sigma}$ and
$n_{j}=n_{j\uparrow}+n_{j\downarrow}$ are the corresponding number
operators and the $SU(2)$ spin operators are given by $S_j^{z}=
\frac{1}{2}(n_{j\uparrow}-n_{j\downarrow})$,
$S_j^-=c^{\dagger}_{j\downarrow} c_{j\uparrow}$ and
$S_j^{+}=c^{\dagger}_{j\uparrow} c_{j\downarrow}$.

The first term in (\ref{hloc1}) is the single-particle hopping, the
second one is known as bond-charge interaction. $U$ and $V$ denote the
on-site and nearest-neighbor Coulomb interaction, respectively. In
addition, we included a $XXZ$-type spin interaction with exchange
constants $J_{xy}$ and $J_z$ between nearest neighbour sites and a
pair-hopping term $Y$.  $\mu$ is the chemical potential and $Z$ the
coordination number of the $d$-dimensional lattice.

The local Hamiltonian can easily be diagonalized. Denoting an empty
site by $0$, a site occupied by an electron with spin $\sigma
=\uparrow,\downarrow$ by $\sigma$, and a doubly occupied site by $2$
we find the following 16 eigenstates and their respective energies:
\begin{eqnarray}
{\rm state} & \qquad & {\rm eigenvalue} \nonumber \\[2mm]
|00\rangle && E_1=\frac{U}{2Z}+V \nonumber \\[2mm]
|{\sigma 0}\rangle  \pm |{0\sigma}\rangle && E_2^{(\pm)}=\mp t
  + \frac{\mu}{Z}  \nonumber \\[2mm]
|{\sigma\sigma}\rangle && E_3 = -\frac{U}{2Z}+\frac{J_z}{4}
  + \frac{2\mu}{Z}  \nonumber \\[2mm]
|{\uparrow\downarrow}\rangle + |{\downarrow\uparrow}\rangle && E_4=
  -\frac{U}{2Z} +\frac{J_{xy}}{2} -\frac{J_z}{4}
  + \frac{2\mu}{Z}\nonumber \\[2mm]
|{20}\rangle-|{02}\rangle&& E_5=\frac{U}{2Z}-V-Y+ \frac{2\mu}{Z}
  \nonumber \\[2mm]
|{\psi_\pm}\rangle=\alpha_\pm\left(|{\uparrow\downarrow}\rangle-
  |{\downarrow\uparrow}\rangle\right)+(|{20}\rangle+|{02}\rangle)
  && E_6^{(\pm)}   \nonumber \\[2mm]
|{\sigma 2}\rangle\pm|{2\sigma}\rangle && E_7^{(\pm)}=\pm (t-2X)
   +\frac{3\mu}{Z}\nonumber \\[2mm]
|{22}\rangle && E_8 = \frac{U}{2Z}+V+ \frac{4\mu}{Z}\nonumber \\[2mm]
\label{locstates}
\end{eqnarray}
with
\begin{equation}
E_6^{(\pm)}={\frac{1}{2}} \left(Y-V-\frac{J_{xy}}{2}-\frac{J_z}{4}+
  \frac{4\mu}{Z} \right)\pm \beta
\end{equation}
and
\begin{equation}
\alpha_\pm = \frac{-\frac{U}{Z}+V-Y-\frac{J_{xy}}{2}
   -\frac{J_z}{4} \pm 2\beta}{4(X-t)}
\end{equation}
where $\beta =\sqrt{\left(\frac{V}{2}-\frac{U}{2Z}
 -\frac{Y}{2}-\frac{J_{xy}}{4} - \frac{J_z}{8}\right)^2+4(X-t)^2}$.

For the special case $t=X$, which will of some importance in the
following, this simplifies to
\begin{eqnarray}
|{\psi_+}\rangle &=& |{\uparrow\downarrow}\rangle
-|{\downarrow\uparrow}\rangle ,
  \nonumber \\[2mm]
|{\psi_-}\rangle &=& |{20}\rangle +|{02}\rangle ,
\end{eqnarray}
with corresponding energies
\begin{eqnarray}
E_6^{(+)}&=& -\frac{U}{2Z} - \frac{J_{xy}}{2}-\frac{J_z}{4}
+\frac{2\mu}{Z}, \nonumber \\[2mm]
E_6^{(-)}&=& \frac{U}{2Z}- V+Y+\frac{2\mu}{Z}.
\end{eqnarray}

First of all we look at the $\eta$-pairing states with momentum $P$,
\begin{equation}
|{\psi_N^{(P)}}\rangle =\left(\eta^{\dagger}_P\right)^N
| \, 0 \, \rangle, \qquad
\eta^{\dagger}_p = \sum_{j=1}^L e^{iPj}c^{\dagger}_{j\downarrow}
c^{\dagger}_{j\uparrow}.
\label{eta}
\end{equation}
As already mentioned these states exhibit ODLRO and are thus
superconducting \cite{pairing}.

The $\eta$-pairing state with $P=0$ is an eigenstate of ${\cal H}$ for
$t=X$, $2V=Y$ \cite{dBKS}. In order to make it an optimal ground state
we first observe that $|{\psi_N^{(0)}}\rangle $ can be built
completely from the local states $|{00}\rangle $, $|{22}\rangle $ and
$|{20}\rangle +|{02}\rangle $. All three states are already local
eigenstates with energy $E_0=E_1=E_6^{(+)}=E_8=U/2Z + V$, if we
choose $\mu=0$. Demanding that $E_0$ is the local ground state
energy one recovers the result obtained in \cite{dBKS} by using
Gerschgorin's theorem:
\begin{eqnarray}
V&\leq & 0, \nonumber \\[2mm]
-\frac{U}{Z}&\geq &\max\Bigl(2|t|+2V, V-\frac{J_z}{4},
V+\frac{|J_{xy}|}{2}+\frac{J_z}{4}\Bigr).
\label{etaineq}
\end{eqnarray}

The state with momentum $\pi$ is an eigenstate of ${\cal H}$ for
$Y=-2V$.  $|{\psi_N^{(\pi)}}\rangle $ is now built from the local
states $|{00}\rangle $, $|{22}\rangle $ and $|{20}\rangle
-|{02}\rangle $. These states have to be made local ground
states. Again for $\mu =0$ they all already have the same energy
$E_0=E_1 =E_5=E_8=U/2Z+V$. All other energies have to be larger.  This
yields the following inequalities:
\begin{eqnarray}
V&\leq & 0, \nonumber \\[2mm]
-\frac{U}{Z}&\geq &\max\Bigl(2|t|+2V, V-\frac{J_z}{4},2|t-2X|+2V,
V-\frac{J_{xy}}{2}+\frac{J_z}{4}, \nonumber \\[2mm]
& &\phantom{\geq \max\Bigl(2}V+\frac{J_{xy}}{2}+\frac{J_z}{4}-
\frac{(t-X)^2}{V}\Bigr).
\label{piineq}
\end{eqnarray}

There exist also $\eta$-pairing states with momentum $P\neq
0,\pi$. These appear for instance as ground states of model
(\ref{hloc1}) where $t=X$ and $U\leq -4t$, but all other interaction
constants are zero \cite{dBKS,AA,as}. These states are eigenstates
only for $t=X$ and $Y=V=0$. It is straightforward to derive a similar
inequality as for the states with momentum $0$ or $\pi$. It is just
(\ref{etaineq}) with $V$ put equal to $0$.

There are other interesting states for which the OGS method can be
applied, e.g.\ the following paramagnetic, N\'eel, charge-density-wave
and ferromagnetic states at half-filling ($N=L$, where
$N=\sum_{j=1}^Ln_{j}$ is the total number of particles),
\begin{eqnarray}
|{para}\rangle &=&\prod_{j\in {\cal A}}c^{\dagger}_{j\uparrow}
  \prod_{j\in {\cal A}'} c^{\dagger}_{j\downarrow}
  | \, 0 \, \rangle , \\[2mm]
|{Neel}\rangle &=&\prod_{j\in {\cal B}}c^{\dagger}_{j\uparrow}
  \prod_{j\in {\cal B}'}c^{\dagger}_{j\downarrow}
  | \, 0 \, \rangle , \\[2mm]
|{CDW}\rangle &=&\prod_{j\in {\cal B}}c^{\dagger}_{j\uparrow}
   c^{\dagger}_{j\downarrow}| \, 0 \, \rangle ,\\[2mm]
|{F}\rangle &=&\prod_{j}c^{\dagger}_{j\uparrow}| \, 0 \, \rangle .
\label{ferro}
\end{eqnarray}
${\cal A}$ and ${\cal A}'$ are arbitrary disjoint sets of lattice
points which together span the whole lattice. The states
$|{Neel}\rangle $ and $|{CDW}\rangle $ are defined on a bipartite
lattice with odd and even sublattices ${\cal B}$ and ${\cal B}'$.

If we want $|{para}\rangle $ to become an optimal ground state we
have to make all the states $|{\sigma \sigma'}\rangle $ with
$\sigma,\sigma'=\uparrow,\downarrow$ local ground states. From
({\ref{locstates}) we see that we have to choose $J_{xy}=J_z=0$ and
$t=X$ in order to get the correct local ground states.  These then
have the energy $E_3=E_4=E_6^{(+)}=-U/2Z+2\mu/Z$. All other
energies have to be higher which finally leads to the condition
\begin{equation}
\frac{U}{Z}\geq \max\left(2|t|+\frac{2|\mu |}{Z},V+|Y|,
-V+\frac{2|\mu |}{Z}\right).
\label{para_ineq}
\end{equation}
For $Y=0$ and $V\geq 0$ this problem has already been investigated in
\cite{strack1}. In this case the bound (\ref{para_ineq}) is better
than the one found in \cite{strack1}. In fact, for $\mu =0$ it is exactly the
improved bound found by Ovchinnikov \cite{ovchin} using the
Gerschgorin approach.

Similarly, for the N\'eel state $|{Neel}\rangle $ we get the
restrictions $t=X$ and $J_{xy}=0$ (from $E_4=E_6^{(+)}$) and the
inequalities
\begin{eqnarray}
J_z&\geq& 0,\nonumber \\[2mm]
\frac{U}{Z}&\geq& \max\left(-V-\frac{J_z}{4}+\frac{2|\mu |}{Z},
2|t|-\frac{J_z}{2}+\frac{2|\mu |}{Z}, V+|Y|-\frac{J_z}{4}\right).
\label{Neel_ineq}
\end{eqnarray}

Turning to the state $|{CDW}\rangle $ we see that this state can be
constructed completely from the local states $|{20}\rangle $ and
$|{02}\rangle $. In order to make these local ground states we have
to choose $t=X$, but also $Y=0$. The local ground state energy is
then $E_5=E_6^{(-)}=U/Z-V+2\mu/Z$. The condition that the remaining
energies are higher leads to the inequalities
\begin{eqnarray}
V&\geq& \frac{|\mu |}{Z},\nonumber \\[2mm]
-\frac{U}{Z}&\geq& \max\left(2|t|-2V+\frac{2|\mu |}{Z},
-V-\frac{J_z}{4},-V+\frac{|J_{xy}|}{2}+\frac{J_z}{4}\right),
\label{CDW_ineq}
\end{eqnarray}
which reduces to Ovchinnikov's improved result \cite{ovchin} for
$J_{xy}=J_z=0$.

The OGS approach shows why a generalization to the non-half-filled
case is difficult.  Away from half-filling also other states (like
$|{\sigma 0}\rangle \pm |{0\sigma}\rangle $) would have to be used in
the construction of the global ground state. Thus the local ground
state degeneracy would have to be larger. In most cases this leads to
too much restrictions on the interaction parameters.

For the fully polarized ferromagnetic state (at half-filling)
$|{F}\rangle $ the local ground state is only $|{\sigma \sigma}\rangle
$. Therefore we do not get restrictions like $J_{xy}=J_z=0$ or $t=X$
in this case.  The local ground state energy is
$E_3=-U/2Z+J_z/4+2\mu/Z$. This is a lower bound for the other local
energies, leading to the inequalities
\begin{eqnarray}
-J_z&\geq& |J_{xy}|,\nonumber \\[2mm]
\frac{U}{Z}&\geq & \max\Biggl(2|t|+\frac{J_z}{2}+\frac{2\mu}{Z},
2|t-2X|+\frac{J_z}{2}-\frac{2\mu}{Z}, \frac{J_z}{4}-V+\frac{2|\mu |}{Z},
V+Y+\frac{J_z}{4},\nonumber \\[2mm]
& &\qquad \phantom{abc}V-Y+\frac{J_z}{4}
-\frac{8(t-X)^2}{J_{xy}+J_z}\Biggr).
\label{ferro_ineq}
\end{eqnarray}
If one is interested in the sector with a fixed particle number $N$,
one can regard the bounds as function of $\mu$ and try to find the value
of $\mu$ which optimizes these bounds.
If a state is a ground state of ${\cal H}$ at some
fixed particle number $N$, then it is also a ground state of ${\cal
H}+\mu N$.  Hence we only need to require that a state is a ground
state for {\it some} value of $\mu$, the result
for arbitrary $\mu$ follows then immediately. Therefore one can
sometimes improve the results of the optimal ground state approach, by
first finding inequalities for an arbitrary value of the chemical
potential, and subsequently optimize with respect to $\mu$.
For example, if we find inequalities $U\geq a+\mu$ and $U\geq b-\mu$,
then the best value of $\mu$ is $(b-a)/2$, and thus the inequality
$U\geq (a+b)/2$.

For (\ref{para_ineq}), (\ref{Neel_ineq}), and (\ref{CDW_ineq}) one
obviously has to choose $\mu=0$. In the case of the ferromagnetic
state the second inequality of (\ref{ferro_ineq}) can be replaced by
\begin{eqnarray}
\frac{U}{Z}&\geq & \max\Biggl(
   |t|+|t-2X|+\frac{J_z}{2}, |t-2X|-\frac{V}{2} +\frac{3J_z}{8},
   |t|-\frac{V}{2}+\frac{3J_z}{8}, \nonumber \\[2mm]
& &\qquad \frac{J_z}{4}-V, V+Y+\frac{J_z}{4},
   V-Y+\frac{J_z}{4}-\frac{8(t-X)^2}{J_{xy}+J_z}\Biggr).
\label{ferro_ineq_improv}
\end{eqnarray}
This bound improves the one found in \cite{strack2}.

We also applied the Gerschgorin approach to the states considered
here.  In all cases the results are not better than the ones obtained
from the OGS approach with optimization with respect to $\mu$
\cite{dbs}.

In summary, we have shown that the OGS approach is -- dispite its
simplicity -- a very useful tool for the construction of exact ground
states.  It allows to reproduce or even improve bounds for stability
regions previously found by different methods
and yields better insights into the physics.
E.g.\ if the
interaction parameters are changed from a situation where the
inequalities are satisfied to one where they are not, the ground
state typically first is non-degenerate or doubly degenerate
(except $|{para}\rangle$, which is always a highly degenerate
 ground state), becomes highly degenerate
at the OGS bounds (where levels of the Hamiltonian cross),
and finally cease to be the exact ground state.  In addition, by
examining which local eigenstates become the new local ground states
one can try to predict some properties of the new ground state once the
bounds are violated. We believe that these bounds are as sharp as
possible. If an inequality is violated we have a new local ground
state which might be used to construct a variational state with an
energy lower than that of the ``optimal'' state.
We will come back to these points in a
future publication \cite{dbs} where we also discuss why the OGS
approach usually yields bounds which are as least as good as those of the
Gerschgorin approach.
\\[4mm]
We like to thank A.\ Kl\"umper, V.\ Korepin, A.\ Montorsi,
E.\ M\"uller-Hartmann and J.\ Zittartz for useful discussions.  JdB is
supported in part by NSF grant No.~PHY-9309888.  AS gratefully
acknowledges financial support by the Deutsche
Forschungsgemeinschaft and through the Sonderforschungsbereich 341. He
also thanks the Institute for Theoretical Physics in Stony Brook,
where this work was started, for its hospitality.
\\[5mm]
\frenchspacing

\end{document}